\documentclass[aps,pra,showpacs]{revtex4-1} 

\usepackage{amssymb}
\usepackage{amsmath}
\usepackage{graphicx}
\usepackage{bm}

\begin{document}

\title{An alternative resolution to the Mansuripur paradox}

\author{Francis Redfern}
\altaffiliation[permanent address: ]{1904 Corona Drive, Austin, Texas 78723}
\affiliation{Texarkana College, Texarkana, TX 75599}
\date{\today}

\begin{abstract}
In 2013 an article published online by the journal \textit{Science} declared
that the
paradox proposed by Masud Mansuripur was resolved. This paradox concerns a point
charge-Amperian magnetic dipole system as seen in a frame of reference where
they are at rest and one in which they are moving. In the latter frame an
electric dipole appears on the magnetic dipole. A torque is then exerted upon
the electric dipole by the point charge, a torque that is not observed in the
at-rest frame. Mansuripur points out this violates the relativity principle and
suggests the Lorentz force responsible for the torque be replaced by the
Einstein-Laub force. The resolution of the paradox reported by
\textit{Science}, based on numerous papers in the physics literature, preserves
the Lorentz force but depends on the concept of hidden momentum. Here I
propose a different resolution based on the overlooked fact that the
charge-magnetic dipole system contains linear and angular electromagnetic field
momentum. The time rate of change of the field angular momentum in the frame
through which the system is moving cancels that due to the charge-electric
dipole interaction. From this point of view hidden momentum is not needed in
the resolution of the paradox.
\end{abstract}


\maketitle

\section{Copyright information}

This is an author-created, un-copyedited version of an article accepted for
publication/published in Physica Scripta. IOP Publishing Ltd is not responsible
for any errors or omissions in this version of the manuscript or any version
derived from it. The Version of Record is available online at 10.1088/0031-8949/91/4/045501.

\section{Introduction}\label{introduction}

In 2012 \cite{Mansur} Mansuripur brought attention to a paradox involving
an Amperian magnetic dipole (a dipole produced by, for example, a current loop)
in the vicinity a point charge. To set up the paradox, a charge and a magnetic
dipole are stationary with respect to each other and arranged such that a
position vector from the charge to the dipole is perpendicular to the magnetic
moment of the dipole. (This is a special case. In general the magnetic
moment must not lie along the line of the position vector.)  Mansuripur pointed
out that in the reference frame of the charge and magnetic dipole, which I will
refer to as the $S'$ frame, there would be no interaction observed between the
magnetic dipole and the charge. However, he also noted that an observer in whose
frame (which I'll call $S$) the charge-dipole arrangement moves in a direction
parallel to the charge-dipole line (or with a velocity component parallel to
that line) would detect a torque acting on the dipole,
a torque not observed in the $S'$ frame. This torque is due to the point charge
acting on an electric dipole that is seen in the $S$ frame by the ``stationary''
observer. This situation violates the principle of relativity; therefore
Mansuripur suggests the Lorentz force that produces the torque should be
replaced by the Einstein-Laub force \cite{Einstein} which does not.

This paradox was addressed earlier by Spavieri \cite{Spav, McD1} and is related
to the paradox posed by Shockley and James \cite{Shock}. Since Mansuripur's
paper was published, there have been numerous papers published addressing this
paradox \cite{Cross,McD1,Ivezic,Mansur2, Mansur3,mit}. Many of these analyses
appeal to the presence of ``hidden momentum'' to explain it
\cite{Cross, McD1,mit,Griff,Cho}; Mansuripur, however, rejects the hidden
momentum explanation in favor of Einstein-Laub \cite{Mansur2,McD1}. A news
article published by the journal
\textit{Science} \cite{Cho} went so far as to declare the paradox resolved, but
this requires hidden momentum to exist in the charge-magnetic dipole system.

A crucial aspect of this problem is the electromagnetic linear and angular field
momentum possessed by the point charge-magnetic dipole combination \cite{Furry}.
I intend to show that an observer in $S$ sees the electromagnetic field angular
momentum of the combination, which is constant in $S'$, changing in time. This
changing angular momentum offsets the problematic torque identified by
Mansuripur such that the total angular momentum of the combination remains
constant. Trying to explain the paradox by appealing to hidden momentum, as many
have done, is not correct from this point of view.

Another important aspect is the
misinterpretation of the electric dipole moment seen in $S$ being due to a
Lorentz contraction effect. The conventional thinking \cite{Rindler} is that the
electric dipole appears on the moving magnetic dipole due to differential
Lorentz contraction of charged particles moving at different speeds as seen by a
stationary observer. Imagine a current loop moving toward you
such that the plane of the loop lies in your line of sight. That side of the
loop where the positive current flows in your direction would appear to be
positively charged, while the opposite side of the loop would appear to be
negatively charged. In the Lorentz-contraction point of view there is an actual
charge separation on the loop: in the $S$ frame one side is \textit{actually}
positive and the other side is \textit{actually} negative. However, this is not
the way the electric dipole really arises.

It arises as a result of the
relativity of simultaneity \cite{Redfern}, not Lorentz contraction. The loop
has no ``\textit{real}'' charge separation, only an apparent one due to
relativistic time differences between the $S$ and $S'$ frames at different
locations on the loop. This understanding is key: If there were an actual
charge separation on the loop, the resulting torque would increase the current
and the magnetic moment, an unphysical effect not seen in the Lorentz
transformation of the electromagnetic fields.

\section{The electromagnetic fields at the point charge in the $S'$ and $S$ frames}

\begin{figure}[ht!]
\label{standard config}
\centering
\includegraphics[width=6in]{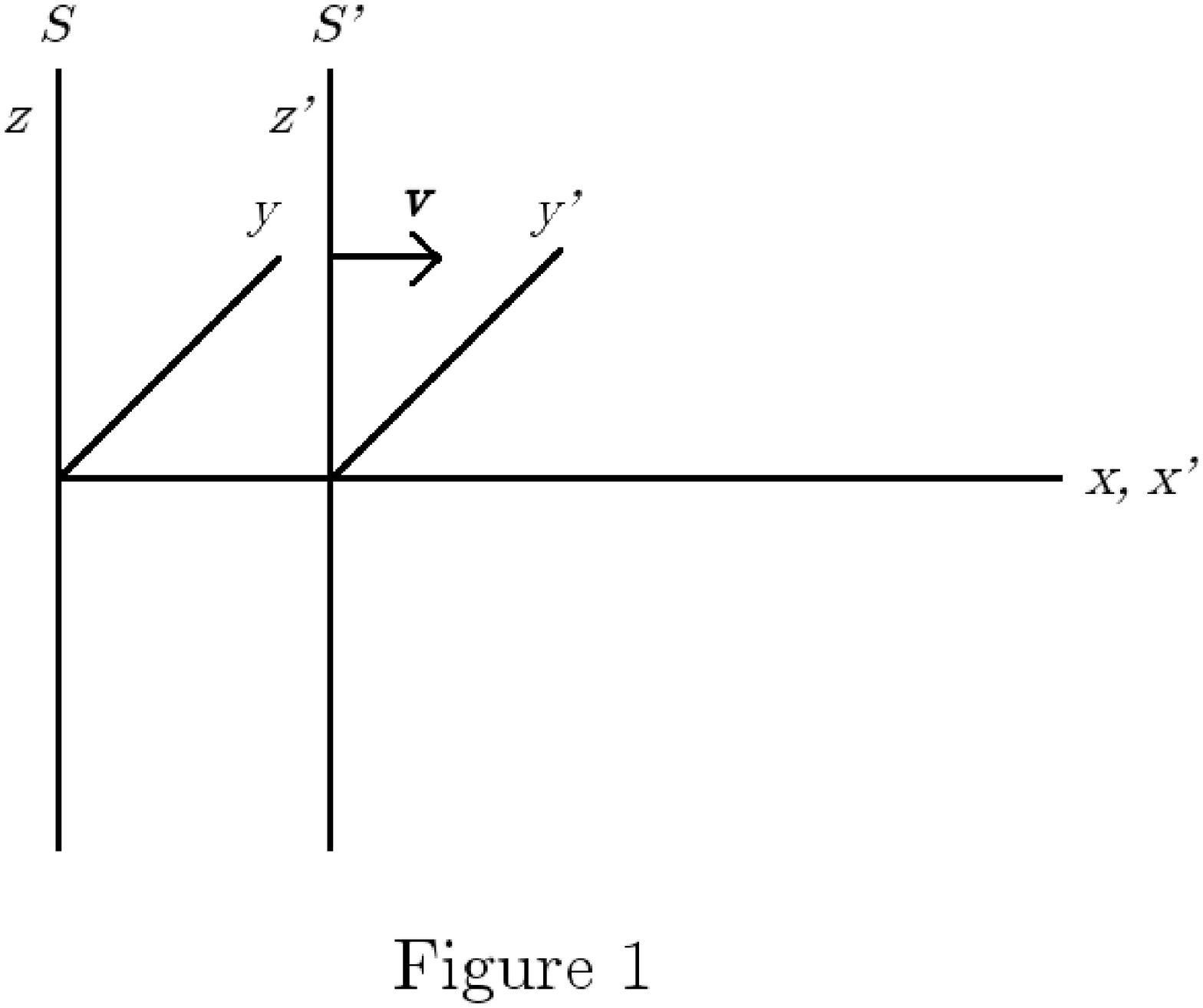}
\caption{Standard configuration of frames $S$ and $S'$. $S'$ is moving in the
positive $x$ direction with speed $v$ where the axes of both frames coincide at
$t = t' = 0$.}
\end{figure}

\begin{figure}[ht!]
\label{t = 0 in S}
\centering
\includegraphics[width=6in]{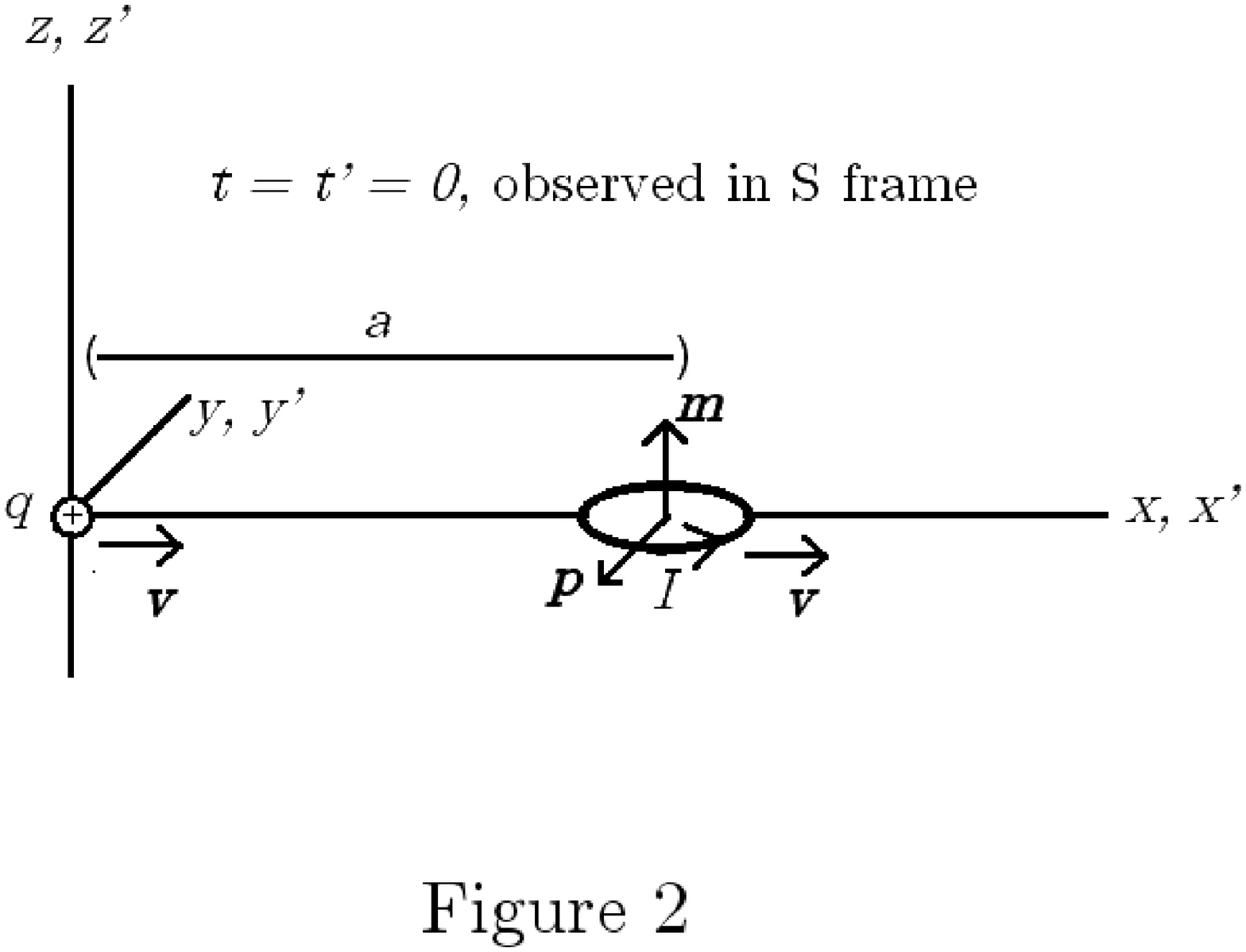}
\caption{The charge-current loop (magnetic dipole) system as seen in frame $S$
at $t = t' = 0$. $\bm{p}$ is the induced electric dipole.}
\end{figure}

The paradox is set up as follows. (See figures 1 and 2.) In frame $S'$ a loop of
current of radius $R$
lies in the $x'$-$y'$ plane with a positive current such that its magnetic
dipole, $\bm{m}'$, is in the $z'$ direction; that is, the positive current
circulates in the positive sense about an axis parallel to the $z'$ axis. A
positive point charge $q$ is located at the origin. The
$S'$ frame, with the charge and current loop, is moving in ``standard
configuration'' \cite{Rindler2}, with speed $v$ in the positive $x$
direction of the $S$ frame and where the space axes of the frames
coincide at $t = t' = 0$. In this configuration, $y = y'$ and $z = z'$.
Since the distance between the center of the loop
and the point charge is $a$ in the $S$ frame, the center of the loop is at
$(\gamma a, 0, 0)$ in the $S'$ frame where $\gamma$ is the Lorentz factor,
\begin{equation}
\label{gamma}
\gamma = \frac{1}{\sqrt{1 - \dfrac{v^2}{c^2}}}.
\end{equation}

In frame $S'$ the loop only produces a magnetic field. Inside and in the
immediate vicinity of the loop, the magnetic field is in the positive $z'$
direction. Everywhere else, the field is in the negative $z'$ direction. At the
position of the charge $q$, the field is, in SI units,
\begin{equation}
\label{Bm' at q}
\bm{B}'_m = -\frac{\mu_o}{4\pi}\frac{m'}{\gamma^3 a^3}\bm{\hat{k}}.
\end{equation}
(From now on the primes will be dropped for quantities that are the same in both
reference frames. Also, some quantities that only appear with reference to the
$S'$ frame, such as the loop radius $R$, will be unprimed.) Lorentz-transforming
the magnetic field from $S'$ to $S$, you get
\begin{equation}
\label{Bm at q}
\bm{B}_m = \gamma\bm{B}'_m;
\end{equation}
that is, the field is increased by a factor of $\gamma$ and is in the same
direction as in $S'$. The electric field at $q$ in the $S$ frame at $t = 0$
resulting from the transformation is
\begin{equation}
\label{Em at q}
\bm{E}_m = -\frac{\mu_o}{4\pi}\frac{m'v}{\gamma^2 a^3}\bm{\hat{j}}.
\end{equation}

The electric field that results from the transformation of the magnetic field
from $S'$ to $S$ does not look like that of an actual electric dipole. For
example, the field has no $x$ component. This can be seen looking at the
equation of the field,
\begin{equation}
\label{induced field}
\bm{E}_m = \frac{\gamma\mu_o m'v}{4\pi r'^3}\left[\frac{3(z^2\bm{\hat{j}}
- yz\bm{\hat{k}})}{r'^2} - \bm{\hat{j}}\right],
\end{equation}
when the magnetic dipole is at the origin and where $r'$ is the Lorentz
transformed distance from the origin.
Also note that since the magnetic moment is constant in the $S'$ frame, it is
also constant in the $S$ frame. This shows that the magnetic moment of the
current loop is not increasing, meaning the troublesome torque of Mansuripur
\cite{Mansur} seen in the $S$ frame is having no effect on the current loop in
that frame.

With an electric field at the position of the charge $q$ in $S$ whereas $q$ is
stationary with respect to the magnetic dipole, it might be expected that the
charge experiences a force in the $S$ frame that it
does not experience in the $S'$ frame. Calculation of the Lorentz force on the
charge shows this is not the case. Since $q$ is moving in the positive $x$
direction in the $S$ frame, its current-density four-vector at $t = 0$ is
\begin{equation}
\label{4J of q in S}
J^{\mu} = \gamma (vq \delta(\bm{r}), 0, 0, cq\delta(\bm{r})),
\end{equation}
where $\delta(\bm{r})$ is the Dirac delta function and $\bm{r}$ is a position
vector centered at the origin of the $S$ frame. The covariant way to
calculate the force is by employing the electromagnetic field tensor
$E^{\mu\nu}$ as follows, using the Einstein summation convention,
\begin{equation}
\label{Lorentz force on q}
E^{\mu\nu}J_{\nu} =
\begin{pmatrix}
0 & B_m & 0 & 0 \\
-B_m & 0 & 0 & -\dfrac{E_m}{c} \\
0 & 0 & 0 & 0 \\
0 & \dfrac{E_m}{c} & 0 & 0 \\
\end{pmatrix}
\begin{pmatrix}
-\gamma vq\delta(\bm{r}) \\ 0 \\ 0 \\ \gamma cq\delta(\bm{r}) \\
\end{pmatrix}
= \begin{pmatrix}
0 \\ \gamma vqB_m\delta(\bm{r}) - \gamma qE_m\delta(\bm{r}) \\ 0 \\ 0 \\
\end{pmatrix}.
\end{equation}
When Eqs. (\ref{Bm at q}) and (\ref{Em at q}) are applied to this result, it is
seen that the force four-vector on the charge $q$ is zero in the $S$ frame as
it must be, since it is zero in the $S'$ frame.

\section{The force density four-vector on the current loop in the $S'$ and $S$
frames}

First I will calculate the force four-vector on the current loop in the $S'$
frame and then Lorentz-transform this force to the $S$ frame. The electric field
at a point $x' = \gamma a + Rcos\phi, y' = Rsin\phi$ on the loop due to the
charge $q$, where $\phi$ is the local azimuth angle measured in the positive
direction from the $x'$ axis, is given by
\begin{equation}
\label{E'}
\bm{E}' = \frac{1}{4\pi\epsilon_o}\frac{q(\gamma\bm{a} + \bm{R})}
{(\gamma^2 a^2 + R^2 + 2\gamma aRcos\phi)^{3/2}},
\end{equation}
where $\bm{a} = a\bm{\hat{i}}$ and $\bm{R} = R(cos\phi\bm{\hat{i}} +
sin\phi\bm{\hat{j}})$. The loop carries a current density given by
\begin{equation}
\label{J' of loop}
J^{\mu'} = \rho u(-sin\phi, cos\phi, 0, 0), \quad \textrm{that is}, \quad
J_{x'} = -\rho u sin\phi \quad \textrm{and} \quad J_{y'} = \rho u cos\phi,
\end{equation}
where $\rho$ is the charge density of the current and $u$ is the drift speed.
(Note that I drop the four-vector notation when referring to specific components
of a four-vector or four-tensor. Contravariant and covariant vectors are the
same in special relativity.) Breaking up the electric field into $x$ and
$y$ components (no $z$ component is present at the loop) and applying the
Lorentz electromagnetic field tensor, you get
\begin{equation}
\label{force calc on loop}
E^{\mu'\nu'}J_{\nu'} =
\begin{pmatrix}
0 & 0 & 0 & \dfrac{E_{x'}}{c} \\
\\
0 & 0 & 0 & \dfrac{E_{y'}}{c} \\
\\
0 & 0 & 0 & 0 \\
\\
-\dfrac{E_{x'}}{c} & -\dfrac{E_{y'}}{c} & 0 & 0 \\
\end{pmatrix}
\begin{pmatrix}
-J_{x'} \\ \\ -J_{y'} \\ \\ 0 \\ \\ 0 \\
\end{pmatrix}
= \begin{pmatrix}
0  \\ \\ 0 \\ \\ 0 \\ \\ \dfrac{J_{x'} E_{x'}}{c} +\dfrac{J_{y'} E_{y'}}{c} \\
\end{pmatrix}
\end{equation}
The force density in the time slot is seen to be
\begin{equation}
\label{f'4 in S'}
f_{ct'} = \frac{J_{x'} E_{x'}}{c} +\frac{J_{y'} E_{y'}}{c}.
\end{equation}
The Lorentz-transformed force density four-vector in the $S$ frame is
\begin{equation}
\label{f4 in S}
f^{\mu} = (\gamma \frac{v}{c}f_{ct'}, 0, 0, \gamma f_{ct'}).
\end{equation}
Assuming the distance $\gamma a$ is much greater than the loop radius $R$, the
electric field components on the loop in $S'$ are approximately (Eq. (\ref{E'}))
\begin{equation}
\label{approx E in S'}
E_{x'} \approx \frac{q(\gamma a + Rcos\phi)}{4\pi\epsilon_o\gamma^3 a^3} \quad
\textrm{and} \quad E_{y'} \approx \frac{qRsin\phi}{4\pi\epsilon_o\gamma^3 a^3}.
\end{equation}
When you substitute $E_{x'}$ and $E_{y'}$ from the above
equations and $J_{x'}$ and $J_{y'}$ from Eq. (\ref{J' of loop}) into Eq.
(\ref{f'4 in S'}) and integrate over the volume, you find that the total
four-force on the loop in $S'$ is zero due to the angular dependence on $\phi$.
Of course, the four force must also be zero in the $S$ frame. Nevertheless this
force is responsible for the appearance of the torque that has been so
troublesome, but this is not a ``real'' torque. Rather, in the model examined
here the spatial torque originates from the Lorentz transformation of a time
component in a torque four-tensor. It will turn out that this torque corresponds
to an increase in the spatial angular momentum of the charge-dipole system,
canceling a decrease identified by a calculation to be presented shortly. In
other words, on the whole there is no torque on the system at all.

\section{The torque four-tensor of the current loop in $S'$ and $S$}

In carrying out the calculations in this section, you
assume the current loop diameter is small compared to its distance from the
point charge. Then, in analogy to the formation of a point electric dipole by
mathematically letting the separation between the charges go to zero as the
magnitudes of the charges go to infinity, you form a point Amperian magnetic
dipole from the current loop by allowing its area $A$ to go to zero while the
current $I$ goes to infinity while holding the product $m' = IA$ constant.

Although the net four-force acting on the loop is zero in the $S'$ frame, the
components of the antisymmetric torque four-tensor, given by the volume
integral
\begin{equation}
\label{torque tensor}
\tau^{\alpha\beta} = \int_V (x^\alpha f^\beta - x^\beta f^\alpha)dV,
\end{equation}
in $S'$ acting on the current
loop do not all turn out to be zero. All components but one pair equal zero due
to the $\phi$ dependence in the volume integrals of the torque density and the
fact that $z = 0$ on the loop. The
non-zero pair (symmetric-antisymmetric partners) are $\tau^{2'4'}$, the
component in row 2 (the $y$ row) and column 4 (the time column) of the tensor
and $\tau^{4'2'} = -\tau^{2'4'}$). The calculation of $\tau^{2'4'}$ is carried
out as follows, taking the origin about the center of the loop for the volume
integration of the torque density,
\begin{equation}
\label{tau'24 a}
\tau^{2'4'} = \int_{V'} (y'f_{ct'} - ct'f_y')dV' =\int_{V'} y'f_{ct'}dV'
= \int_{V'}(Rsin\phi)\left(\frac{J_{x'} E_{x'}}{c} +\frac{J_{y'} E_{y'}}{c}\right)dV'.
\end{equation}
To perform the volume integration, you assume that the wire of the loop is
one-dimensional, which lets you make the substitution $\rho dV' = \lambda R
d\phi$ where $\lambda$ is the linear charge density of the charge carriers
responsible for the current. This allows you to write the integral as
\begin{equation}
\label{tau'24 b}
\tau^{2'4'} = \frac{R^2\lambda u}{c}\int_0^{2\pi}(-E_{x'} sin^2\phi +
E_{y'} sin\phi cos\phi)d\phi.
\end{equation}
The second integrand gives zero when integrated over $\phi$. The first integrand
gives
\begin{equation}
\label{tau'24 c}
\tau^{2'4'} = \frac{R^2\lambda u}{c}\int_0^{2\pi}\left(-\frac{q(\gamma a + Rcos\phi)}{4\pi\epsilon_o\gamma^3 a^3}\right)sin^2\phi d\phi
= -\frac{\lambda(u/c)q\pi R^2}{4\pi\epsilon_o\gamma^2 a^2}.
\end{equation}
This torque, when transformed to the $S$ frame, gives rise to a torque about the
$z$ axis, as follows,
\begin{eqnarray}
\label{tau_z}
\tau_z &=& \tau^{12} = \gamma\frac{v}{c}\tau^{4'2'} = \gamma\frac{v}{c}(-\tau^{2'4'}) \nonumber \\
&=& \frac{v}{c^2}\frac{qm}{4\pi\epsilon_o a^2},
\end{eqnarray}
where the magnetic moment in the $S$ frame is $m = \gamma m'$, $m' = I\pi R^2$,
and $I = \lambda u$. This is the torque, pointed out by Mansuripur
\cite{Mansur}, that is central to the paradox.

\section{The time rate of change of the charge-dipole field angular momentum}

According to Furry \cite{Furry} the electromagnetic field angular and linear
momentum associated with a point charge in the vicinity of a magnetic dipole in
the $S'$ frame (assuming the same configuration as previously) are,
respectively, using the symbols in this paper,
\begin{equation}
\label{ang and lin mo}
\bm{L}' = \frac{1}{c^2}\frac{qm'}{4\pi\epsilon_o\gamma a}\bm{\hat{k}} \quad
\textrm{and} \quad \bm{P}' = - \frac{1}{c^2}\frac{qm'}{4\pi\epsilon_o\gamma^2 a^2}\bm{\hat{j}},
\end{equation}
where the angular momentum is taken about the center of the magnet.
His result for angular momentum needed no particular model for the magnet;
however the result for linear momentum was derived using a spherical magnet
model. It is nevertheless appropriate here for a current loop as seen by the
following argument.  Were the magnetic moment to decay to zero, the linear
momentum stored would be entirely transmitted to the point charge
\cite{Redfern2}, converting both field momenta into mechanical momenta. Hence,
it is reasonable that the linear momentum  should satisfy the following
equation.
\begin{equation}
\label{Furry linear momentum}
\bm{L}' = -\gamma a\bm{\hat{i}}\times\bm{P}'.
\end{equation}
Comparison of this equation with $\bm{L}'$ in Eq. (\ref{ang and lin mo})
confirms that the linear momentum found by Furry is correct for this situation.

Thus the antisymmetric angular momentum tensor for the system has non-zero
components $L^{1'2'}$, $L^{2'4'}$ (and their antisymmetric partners) given by
\begin{equation}
\label{L' tensor components}
L_{z'} = L^{1'2'} = -L^{2'1'} = \frac{1}{c^2}\frac{qm'}{4\pi\epsilon_o\gamma a}
\quad \textrm{and} \quad L^{2'4'} = -L^{4'2'} = -ct'P_{y'},
\end{equation}
where $P_{y'}$ is the linear momentum in the $y'$ direction, given by
\begin{equation}
\label{P'}
P_{y'} = -\frac{1}{c^2}\frac{qm'}{4\pi\epsilon_o\gamma^2 a^2}.
\end{equation}
Note that the time component of the angular momentum depends on time, although
the spatial angular momentum vector $\bm{L}_z$ is constant in time. Transforming
the tensor
\begin{equation}
\label{L'}
L^{\mu'\nu'} =
\begin{pmatrix}
0 & L_{z'} & 0 & 0 \\
-L_{z'} & 0 & 0 & -ct'P_{y'} \\
0 & 0 & 0 & 0 \\
0 & ct'P_{y'} & 0 & 0 \\
\end{pmatrix}
\end{equation}
 to the $S$ frame gives the component $L_z = L^{12}$ as follows,
\begin{equation}
\label{L_z}
L_z = \gamma L^{1'2'} + \gamma\frac{v}{c}ct'P_{y'}
= \gamma(L^{1'2'} + vt'P').
\end{equation}
The time derivative of this gives the time rate of change of the angular
momentum, which is the torque involved. Recalling that $t = \gamma t'$, this
torque is
\begin{equation}
\label{dL/dt}
\frac{dL_z}{dt} = vP' = -\frac{v}{c^2}\frac{qm}{4\pi\epsilon_o a^2},
\end{equation}
where the relationship $m = \gamma m'$ has been used. This torque is equal and
opposite to that found in Eq. (\ref{tau_z}). Thus these torques cancel and the
total angular momentum of the system is constant.

\section{Discussion and Conclusion}

I believe some investigations into this paradox have been misled by the concept
that a system containing electromagnetic momentum at rest in a certain
reference frame must necessarily contain some sort of mysterious ``hidden
momentum'' in that frame. (See, for example, \cite{Cross} and \cite{McD1}.)
However, unless you believe in some sort of magical appearance of a
charge-magnetic dipole system out of the blue, there had to be some sort of
assembly of this system -- an interaction between the components of the system,
originally with zero energy and zero momentum, and their environment to put
together a system that has energy and momentum.

Here is one way the system of
Mansuripur's paradox could be put together. Assume the magnet is
given by the model of Shockley and James \cite{Shock} but have its rigid,
charged disks not rotating initially. Now have an external agent bring a point
charge from a distance into its vicinity. No net force is required to do this.
The external agent then applies torque, equal and opposite to the two disks, to
produce rotation, current, and a magnetic dipole, but no net mechanical
angular momentum. The magnetic field will increase in the negative $z$ direction
at the location of the point charge $q$, producing an impulse on the charge
given by $-\bm{P}'$ in Eq. (\ref{ang and lin mo}) \cite{Redfern2,Shock}. To keep
the charge stationary, an equal and opposite impulse must be supplied to the
charge by the external agent. This impulse will store both linear and angular
momentum in the electromagnetic field of the charge-dipole system, and the
external agent will receive an equal and opposite amount of momentum. The
total momentum is zero, but the charge-magnetic dipole does have energy and
electromagnetic field momentum in its rest frame. If you like, the ``hidden
momentum'' can be thought of as residing in the external agent.

From the result of the last section, there is no role for any hidden momentum
in the charge-dipole system to account for the troublesome torque in $S$, as it
is accounted for by the
time rate of change of the transformed field angular momentum in that frame.
Instead of appealing to hidden momentum, it seems preferable to interpret the
paradox as follows. When you calculate the four-force acting on the current loop
in its rest frame due to the presence of the point charge, there is a non-zero
time component which gives rise to a time-component in the torque four-tensor.
When that tensor is transformed to the frame through which the charge and loop
are moving, it results in a non-zero torque, the troublesome
torque of Mansuripur. However, it is also true that the charge-loop combination
has linear and angular field momentum, giving rise to a time dependent component
in a time slot of its angular momentum four-tensor. When the four-tensor is
transformed to the frame through which the combination is moving, there results
a time-dependent spatial angular momentum vector. The time derivative of this
produces a torque equal and opposite to the one arising from the interaction of
the point charge with the current loop. Hence the spatial angular momentum
vector of the system as a whole is constant and the net torque is zero in both
frames of reference.

It should also be emphasized that there is no ``real'' charge
separation on the current loop as seen in a frame through which it is moving. An
actual charge separation on the loop would result in a dipolar electric field
pattern around the loop, but that is not the case. Rather,
the appearance of charge separation is due to the relativity of simultaneity
\cite{Redfern} not differential Lorentz contraction. Hence, there is no
rotational acceleration associated with the torque and no change in the loop's
magnetic moment. This would not be true were one side of the loop actually
positive and the other negative. Were this the case it would be difficult to
understand how the torque due to the point charge was negated by the
electromagnetic field torque.

Finally, I should point out that nothing in this work says anything about the
physical validity of the Lorentz versus the Einstein-Laub formalisms for a
classical Amperian magnetic dipole. Mansuripur has pointed out that when an
electromagnetic system is
treated as a whole both formalisms give the same total force and torque
\cite{Mansur4}. Mansuripur's argument is that the Lorentz formalism needs
to be corrected by including the angular momentum density of the magnetization
of the magnetic dipole that results from Einstein-Laub,
$\bm{r}\times(\epsilon_o \bm{E}\times\bm{M})$, where $\bm{M}$ is the
magnetization. This term cancels the supposed hidden angular momentum density
to resolve the paradox \cite{Mansur4}. As it turns out, this term ``fixes'' the
problem by accounting for the lack of consideration of the role of the field
angular momentum when the Lorentz formalism is not properly applied. Hence,
there is no paradox in either the Lorentz or Einstein-Laub formalisms if the
magnetic dipole is Amperian. Hidden momentum is not needed in either.

\bibliographystyle{apsrev4-1}
\bibliography{mansuripur}

\end{document}